# Is it always worthwhile to resolve the governing equations of plate theories for graded porosity along the thickness?

S.K. Jalali [a*], M.J. Beigrezaee [b], Nicola M. Pugno [a,c]

[a] Laboratory of Bio-Inspired, Bionic, Nano, Meta Materials & Mechanics,

Department of Civil, Environmental and Mechanical Engineering, Università di Trento, via Mesiano, 77, I-38123 Trento, Italy

[b] Department of Mechanical Engineering, Kermanshah University of Technology, Kermanshah, Iran

[c] Griffith Theory Centenary Lab, School of Engineering and Materials Science, Queen Mary University of London, Mile End Road, London E1 4NS, UK

* corresponding author: seyed.kamal.jalali@gmail.com

## Abstract

Functionally graded porous (FGP) plates have been introduced as modern structural members which open a new window to optimal and functional designs. Despite the need to study the effect of graded porosity on the mechanical behavior of FGP plates, it is necessary to consider the very extensive and valuable literature in this field, presenting remarkable closed-form solutions. Hence, this paper aims to answer where is possible to implement the available exact solutions for the analysis of FGP plates. As the special distinction of FGP plates, graded porosity, is reflected in their stiffnesses and moments of inertia coefficients, 12 different functionality of porosity distribution along the thickness are considered and a set of explicit formulation for evaluating these coefficients are presented to be substituted in already provided analytical solutions. Many examples including bending and free vibration of thin and thick FGP plates are exhibited and the influence of the type of porosity distribution is discussed in details. This work can be considered as a guideline for designers to evaluate the effect of graded porosity based on the cornerstone of the huge number of solutions in the precious literature of plate theories.

## 1. Introduction

The application of the plate theory is useful to design related structures and its applications have been extending in the various industries. Plates with various geometrical shapes e.g. rectangular, circular, annular, sectorial, triangular, trapezoidal, skew, polygonal, etc. have been extensively studied in order to evaluate the mechanical behavior of them under various loading conditions such as bending [1], buckling [2], and vibrations [3] even in thermal environments [4]. In addition, all the possible boundary conditions such as clamped, simply supported, free, and elastic foundations have been considered for analyzing the plates. A plate is defined as a three-dimensional mechanical component which one of its dimensions is much smaller than the two other ones. This especial planar geometry is considered to develop some simplified two-dimensional theories, which have been progressive throughout the time, instead of the more complex and general three-dimensional elasticity. These plate theories can be classified into two main categories i.e. the Classical Plate Theory (CPT) first introduced by Kirchhoff [5] and several shear deformation plate theories. The key difference between them is that the former does not take to account the shear stresses along the thickness of the plate while the latter considers it in various assumptions introducing many different formulations for shear deformation of plates. It is shown that only for enough thin plates the results obtained by CPT are reliable while for more than moderately thick plates shear deformation theories need to be adopted. The simplest one, first mentioned by Mindlin and Reissner, is known as the first-order shear deformation plate theory (FSDT) considering constant shear stress along the thickness by defining a shear correction factor. The higher-order shear deformations lead to subsequent plate theories, which the review of them was reported by Reddy [6–8]. These plate theories are widely implemented in the literature [1,6,9–11]. In the new era of composites, e.g. fiber-reinforced laminated composites, functionally graded materials, nanocomposites, porous materials, or smart materials, the plates made of these modern materials were manufactured for achieving the concept of efficient and smart structures. Accordingly, in the last decades, the theories of plates have been upgraded and the associated governing equations been resolved by the researchers in the field for these new applications.

Plates made of fiber-reinforced laminated composites as orthotropic materials, have extensive usage especially in lightweight applications [12]. A comprehensive investigation on the mechanics of laminated composite plates and shells is presented in the book by Reddy [6]. As another composite, functionally graded materials (FGMs) in the classification of isotropic but nonhomogeneous materials have been widely taken into consideration in the aerospace, biomedical, and automotive industries. FGM plates are generally made of two materials whose properties gradually change from one to the other via a function of the position. The most recognized types of FGM plates are made of the mixture of metals and ceramics, mainly as thermal shielding where the metal guarantees the sufficient toughness whereas the ceramic can resistance the vast of temperature changes [13]. For this application the properties of FGM plates need to vary along the thickness direction [14–16], however, the variation along the in-plane directions are also reported [17–19]. A huge number of researches addressed the mechanical behavior of FGM plates by means of either theoretical or numerical solutions [20,21]. Static bending analysis of the FGM circular or rectangular plates under different loading and boundary conditions has been investigated in detail [22–25]. Besides, vibration analysis of the FGM plates have been examined especially due to its application in aerospace applications [26–28]. The fabrication of FGM plates may include some inevitable porosities during the manufacturing process which should be considered as voids for evaluating the mechanical behavior. For assessing the influence of these defects, both randomly or gradually distribution of voids are assumed for studying both FGM plates [29–31] or beams [32–34].

On the other hand, impressive growth of digital manufacturing especially additive manufacturing for mass production of structural members as well as developments in the fabrication of the porous metal foams provide a new opportunity for engineers to utilize the high porous materials in order to achieve optimal and functional designs [35]. Thanks to developing the industrial 3D printers with the possibility of free design of microstructures, it is possible to ideally fabricate any kind of porosity even plates with functionally graded porosity in the desired directions. The application of such functionally graded porous (FGP) plates can be extended in the industries especially where the weight to strength ratio is a critical parameter such as in aircraft and marine technologies; However, few studies have been accomplished in this field. Next, some of researches of FGP have been summarized.

Heshmati and Daneshmand [36] exposed the benefits of using the FGP plates compared to the traditional ones. They analyzed 3D free vibration of FGP plates resting on two-parameter elastic foundations and a semi-analytical approach was presented. The effect of radially graded porosity on the free vibration behavior of thick circular and annular sandwich plates was investigated by Heshmati and Jalali [37]. Both clamped and simply supported boundary conditions were considered and equations of motions were numerically solved by the pseudo-spectral method. Using FSDT, the linear free vibration properties of rectangular porous plates located between two layers of piezoelectric have been investigated by Askari et al. [37]. The results revealed that the type of porosity distribution could significantly affect the vibrational behavior. In addition to FSDT, these authors [38] evaluated the same problem by means of third-order shear deformation plate theory. In another study, CPT has been employed alongside the Von Kármán strain-displacement relationship and stress function in order to evaluate a graphene platelet reinforced sandwich by Li et al. [39]. They computed the dynamic buckling and non-linear vibration of the plate and discussed the influence of Winkler-Pasternak elastic foundation. Damping and thermal environment effects were also taken into account. Rezaei and Saidi studied vibrational [40] and buckling [41] analysis of the moderately thick fluid-infiltrated porous annular sectorial plates using FSDT. Simply supported boundary condition was set for radial edges and the porous network was saturated by the fluid. Their outcomes ascertained that the existence of the fluid in the FGP could increase the fundamental frequency and the critical buckling load. In a similar study, Kamranfard et al. [42] examined the FGP plates with annular sectorial shape under in-plane uniform compressive loading in which buckling and vibration analysis were taken into consideration by means of FSDT. Likewise, simply supported boundary condition was used for radial edges. The main results deduced that increasing the in-plane load leads to decreasing the natural frequency, and the mode shapes of vibration depended on the geometrical parameters and loading values. Vibration and buckling response of the FGP plates were also studied by Yang et al. [28]. They developed FSDT to determine the response of FGP nanocomposite plates reinforced with graphene platelets. Moreover, the displacement field was interpreted by utilizing the Chebyshev-Ritz method. As well as the other research results, they demonstrated that the increase of porosity coefficients caused the decrease of the fundamental natural frequency and the buckling loads. Circular tapered FGP plates were analyzed by Jalali and Heshmati [28]. Unlike the previous studies in which the porosity distribution was in the thickness direction, they introduced three different types of porosity distribution along the radial direction. By means of FSDT, vibration behavior of two different boundary conditions, simply supported or clamped end ones were illustrated alongside the pseudo-spectral method to solve the equations of motions. In addition to FGP plates, researchers have also applying FGP for beams, panels, and shells. Chen et al. [43] assumed two different distributions for porosity in the thickness direction and solved the static bending and elastic buckling problems by using the Timoshenko beam theory. Not only the static problems but also the dynamic analysis of the beam was also investigated. Heshmati and Daneshmand [44] analyzed the vibration response of the two sides clamped FGP beams by using the Timoshenko beam theory. The behavior of vibration of FGP doubly-curved panels and shells of revolution was determined by Zhao et al. [45]. Like most of the researches, they considered the porosity distribution in the thickness direction and general boundary condition was obtained via the modified Fourier series. Finally, they reported the influence of the boundary conditions and parameters of material on free vibration behavior.

FGP plates are novel structural members with many potential applications in a wide variety of industries. Investigation on their responses under different kinds of loading and boundary conditions for various geometrical shapes can open a new window for engineers to achieve optimal designs. Despite some recent studies, the problem is open for many other conditions. On the other hand, a huge number of analytical solutions are provided in the literature for the general problem of isotropic, anisotropic, homogeneous, and nonhomogeneous plates under various types of loadings based on well-known plate theories. The key point of the present paper is to answer this question: Is it always worthwhile to resolve the governing equations of plate theories for graded porosity along the thickness? It is shown that the answer is no, in general. This paper aims to illuminate the way for implementation of analytical solutions in the literature for evaluation the mechanical response of FGP plates by introducing a set of explicit equations for calculating the stiffnesses and moments of inertia of FGP plates having various functionality of porosity distributions. It prevents resolving the governing equations of motions for the FGP plates in the situations solved in the past. In order to validate, various examples are presented to predict the behavior of the FGP plates with different types of porosity distributions along the thickness direction and the obtained results are compared to those from FEM. This paper can be an applicable guideline for designers and engineers to assess the response of the FGP plate alongside the extensive solutions in the literature.

## 2. Problem description

### 2.1. Functionally graded porosity

At first, the relationship between the amount of porosity and density of a porous material should be defined. The porosity can be quantified by introducing the porosity parameter $p$ which varies between zero and one. The density $\rho$ of a porous material is simply related to this parameter as follows:

$$\rho = \bar{\rho}(1 - p) \tag{1}$$

where $\bar{\rho}$ is the density of the bulk material without any porosity ($p = 0$). Because of distributions of pores through the material, the Young's modulus also changes with respect to the value of the bulk material, $\bar{E}$. In this work, an FGP plate of open-cell porosity has been selected because of its capability for achieving high porosity values as well as its possibility for 3D printing manufacturing. For open-cell porous materials, the power law of the power of two can describe the relationship between the Young's modulus and the density accurately [43,46–48]:

$$\frac{E}{\bar{E}} = \left(\frac{\rho}{\bar{\rho}}\right)^2, E = \bar{E}(1 - p)^2 \tag{2}$$

According to Fig.1, consider a plate of the thickness of $h$ where $z$ axis aligned with the thickness direction and the origin of the coordinate locates on the mid-plane of the plate. This plate is made of bulk material of density of $\bar{\rho}$, Young's modulus of $\bar{E}$ and Poisson's ratio of $\bar{\nu}$. For FGP plates with porosity variation along the thickness, the amount of porosity is a function of $z$-coordinate as $p(z)$. If the parameter of $p_m$ represent the maximum porosity in the plate, Eq. (1) can be rewritten as:

$$\begin{aligned} \rho(z) &= \bar{\rho}\big(1 - p(z)\big) \\ p(z) &= p_m F(z) \end{aligned} \tag{3}$$

in which the function $0< F(z)<1$ controls how the porosity varies. As the power law, Eq. (2), is valid for all the points of the porous plate, thus Young's modulus varies in the thickness direction depended on the $z$-coordinate as follows:

$$E(z) = \bar{E}\big(1 - p(z)\big)^2 \tag{4}$$

Three different types of porosity variation along the thickness, named Pyramid (P), Sandglass (S), and Diamond (D) have been assumed which are presented in Fig. 1. For the P-type, the maximum porosity locates at the top of the plate ($z = +0.5h$) while the minimum locates at its bottom ($z = -0.5h$). For the S-type, the maximum and the minimum porosities locate at the mid-plane of the plate ($z = 0$) and at its surfaces ($z = \pm 0.5h$), respectively, while for the D-type is vice versa. It is mentioned that unlike the P-type, the material properties are symmetric with respect to the mid-plane for both the D- and S-types.

Although the porosity variation can be defined as any mathematical function, four common functions with a vast range of varieties are considered, named Linear, Parabolic, Cubic, and Cosine. Consequently, the function of porosity variation along the thickness, $F(z)$, has 12 expressions as:

***Polynomial Function (k=1: Linear, k=2: Parabolic, k=3: Cubic):***

P-type: $F(z) = \left(\frac{1}{2} - \frac{z}{h}\right)^k$

S-type: $F(z) = \left|\frac{2z}{h}\right|^k$ (5a)

D-type: $F(z) = 1 - \left|\frac{2z}{h}\right|^k$

**Cosine Function:**

P-type: $F(z) = \cos(\frac{\pi z}{2h} + \frac{\pi}{4})$

S-type: $F(z) = \left(1 - \cos\left(\frac{\pi z}{h}\right)\right)$ (5b)

D-type: $F(z) = \cos\left(\frac{\pi z}{h}\right)$

In order to refer to each of these porosity distributions, Table 1 names them. Fig 2 shows the variety of the $F(z)$ function through the thickness direction for all the porosity distributions.

**Table 1:** Various types of porosity distributions.

| | **Linear** | **Parabolic** | **Cubic** | **Cosine** |
|---|---|---|---|---|
| **Pyramid (P-type)** | P1 | P2 | P3 | PC |
| **Sandglass (S-type)** | S1 | S2 | S3 | SC |
| **Diamond (D-type)** | D1 | D2 | D3 | DC |

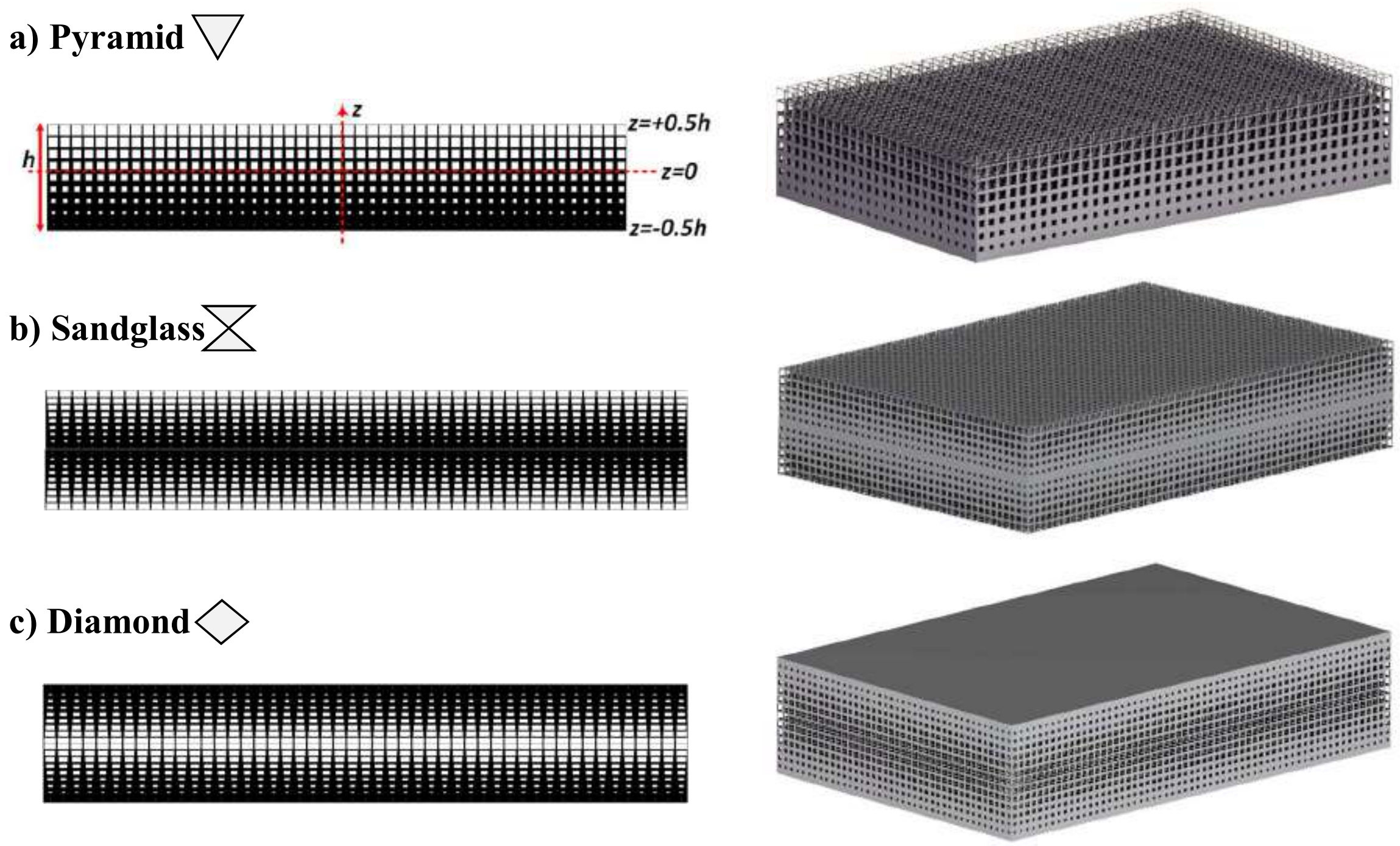

**Fig 1.** Types of variation of porosity along the thickness a) Pyramid, b) Sandglass, c) Diamond.

## 2.2. Stiffnesses and Moments of Inertia

In the governing equations of motions based on the well-known theories of plates, the stiffnesses and the moments of inertia coefficients express the material/structural properties. Thus, the difference between solutions due to changing the material properties of plates is reflected in these coefficients which can be obtained as below:

$$(A, B, D) = \int_{-h/2}^{+h/2} Q(z)(1, z, z^2)dz \tag{6a}$$

where $Q(z) = \frac{E(z)}{1-\bar{\nu}^2}$ (6b)

$$(I_0, I_1, I_2) = \int_{-h/2}^{+h/2} \rho(z)(1, z, z^2)dz \tag{6c}$$

where $A$, $B$ and $D$ are the stretching stiffness, the stretching-bending stiffness, and the bending stiffness, respectively. Moreover, the moments of inertia are represented by $I_0$, $I_1$ and $I_2$. The effect of porosity on the Poisson's ratio has been neglected in this research so that the Poisson's ratio has a constant value of $\bar{\nu}$. It is worth mentioning that for the case of symmetric material respect to the mid-plane occurs in S- and D-types, the function under integral for evaluating two coefficients of $B$ and $I_1$ is becoming odd which result in zero. In addition, if the plate thickness is constant and the mechanical properties change only along the thickness direction, these stiffnesses and moments of inertia coefficients appear as constants in the governing equations of the plate. Otherwise, in the case of either variable thickness plates or variation of porosity along the in-plane axes, the coefficients of stiffnesses and moments of inertia depend also on the in-plane position.

Based on Eq. (6) the coefficients of stiffnesses and moment of inertia for an FGP plate depend on the plate thickness, $h$, density and Young's modulus of the bulk material, $\bar{\rho}$ and $\bar{E}$, and absolutely the maximum value of the porosity $p_m$. To generalize the results, the dimensionless stiffnesses and moments of inertia coefficients are defined as:

$$A^* = \frac{A}{\bar{A}}, B^* = \frac{B}{\bar{A}h}, D^* = \frac{D}{\bar{D}}, I_0^* = \frac{I_0}{\bar{I}_0}, I_1^* = \frac{I_1}{\bar{I}_0 h}, I_2^* = \frac{I_2}{\bar{I}_2} \tag{7a}$$

$$\bar{A} = \frac{\bar{E}h}{(1-\nu^2)}, \bar{D} = \frac{\bar{E}h^3}{12(1-\nu^2)}, \bar{I}_0 = \bar{\rho}h, \bar{I}_2 = \frac{\bar{\rho}h^3}{12} \tag{7b}$$

These dimensionless parameters are only function of the maximum porosity $p_m$ and are independent of the geometric and material parameters. Substituting Eq. (7) into Eq. (6) the explicit formulation for calculating the dimensionless stiffnesses and the moments of inertia coefficients is obtained. It is demonstrated that the relationship between the dimensionless stiffnesses ($A^*$, $B^*$, $D^*$) and the maximum porosity parameter, $p_m$, is in the general form of a second-order polynomial equation, Eq. (8). The coefficients of $a_1$ and $a_2$ are illustrated in Table 2 for different porosity distributions. In addition, the $a_3$ coefficient for evaluating $A^*$and $D^*$ is equal to one and for $B^*$ is equal to zero.

$$\begin{aligned}&(A^*, B^*, D^*) = a_1 p_m^2 + a_2 p_m + a_3,\\ &for\ A^*\ and\ D^*: a_3 = 1, for\ B^*: a_3 = 0\end{aligned} \tag{8}$$

The relationship between the dimensionless moments of inertia ($I_0^*$, $I_1^*$ and $I_2^*$) and the maximum porosity parameter, $p_m$, is given in the general linear form of Eq. (9) whose slope, $b_1$, is reported in Table 2 for different porosity distributions. The $b_2$ coefficient for evaluating $I_0^*$ and $I_2^*$ is equal to one and for $I_1^*$ is equal to zero.

$$\begin{aligned}&(I_0^*, I_1^*, I_2^*) = b_1 p_m + b_2\\ &for\ I_0^*\ and\ I_2^*: b_2 = 1, for\ I_1^*: b_2 = 0\end{aligned} \tag{9}$$

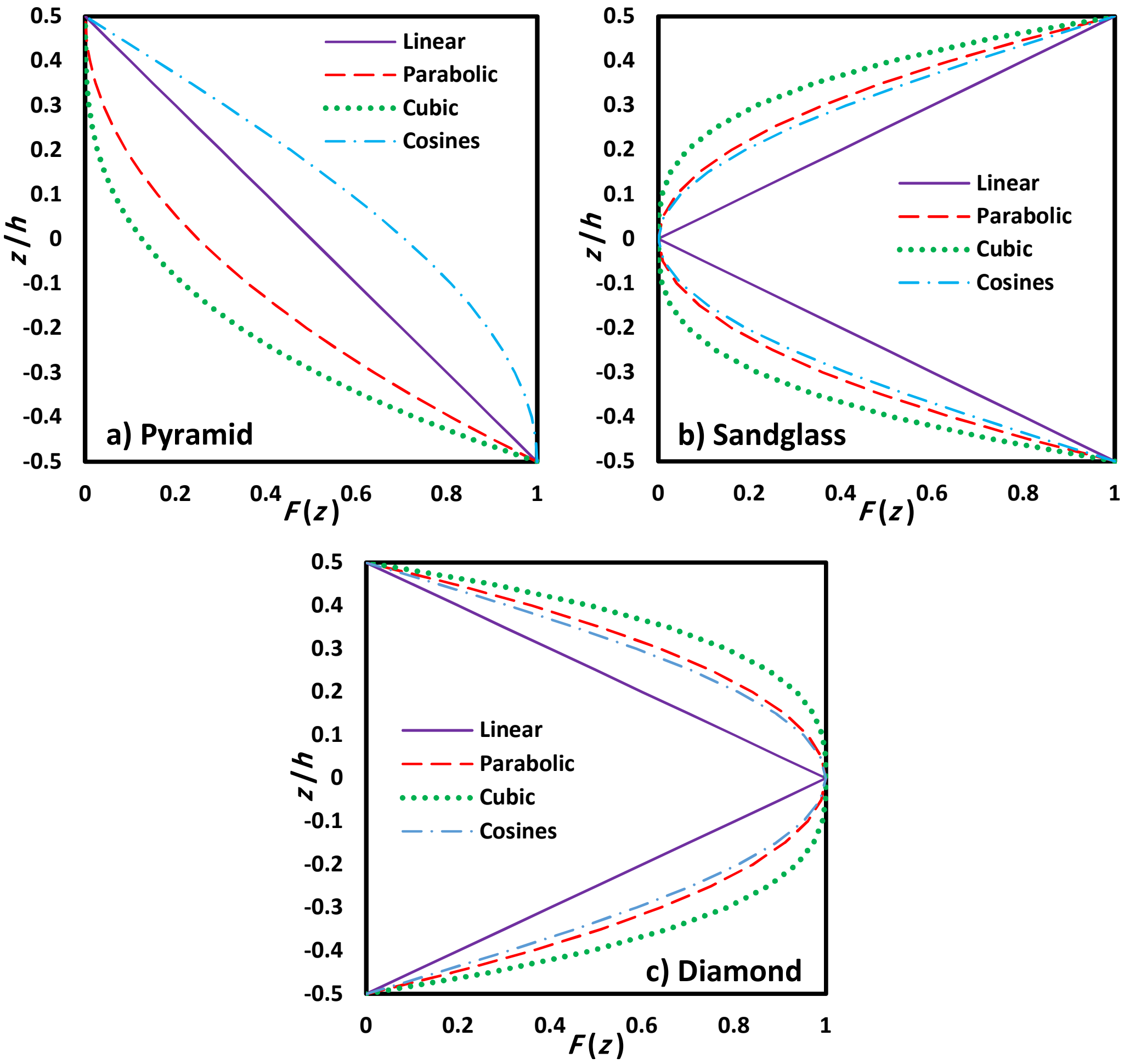

**Fig 2.** Variation of function *F*(*z*) along the thickness of plate. a) Pyramid, b) Sandglass, c) Diamond.

Obviously, for all 12 types of porosity distributions, by increasing the porosity, both the dimensionless stiffnesses and the moments of inertia decreases because increasing the porosity leads to decreasing the stiffness, however, in Eq. (10a-f), the values of dimensionless stiffnesses and moments of inertia for all types of porosity distributions along the thickness are compared. This comparison may be useful as an initial design idea for selection of the proper porosity distribution.

$$A^* \qquad D3 < D2 < PC = DC < P1 = S1 = D1 < SC < P2 = S2 < P3 = S3 \tag{10a}$$

$$B^* \qquad PC < P1 < P3 < P2 \tag{10b}$$

$$D^* \qquad S1 < SC < S2 < PC < S3 = D3 < P1 < D2 < DC < P2 < P3 < D1 \tag{10c}$$

$$I_0^* \qquad D3 < D2 < PC = DC < P1 = S1 = D1 < SC < P2 = S2 < P3 = S3 \tag{10d}$$

$$I_1^* \qquad P3 < P1 = P2 < PC \tag{10e}$$

$$I_2^* \qquad S1 < SC < S2 < PC < S3 = P1 = D3 < P2 = D2 < DC < P3 < D1 \tag{10f}$$

**Table 2:** Coefficients for calculation of the dimensionless stiffnesses and moments of inertia of FGP plates.

| | | Linear | | | Parabolic | | | Cubic | | | Cosine | | |
|---|---|---|---|---|---|---|---|---|---|---|---|---|---|
| | | **P1** | **S1** | **D1** | **P2** | **S2** | **D2** | **P3** | **S3** | **D3** | **PC** | **SC** | **DC** |
| $\boldsymbol{A}^*$ | $a_1$ | 1/3 | 1/3 | 1/3 | 1/5 | 1/5 | 8/15 | 1/7 | 1/7 | 9/14 | 1/2 | $\frac{-8+3\pi}{2\pi}$ | 1/2 |
| | $a_2$ | -1 | -1 | -1 | -2/3 | -2/3 | -4/3 | -1/2 | -1/2 | -3/2 | $-4/\pi$ | $\frac{2(2-\pi)}{\pi}$ | $-4/\pi$ |
| $\boldsymbol{B}^*$ | $a_1$ | -1/12 | 0 | 0 | -1/15 | 0 | 0 | -3/56 | 0 | 0 | $-1/\pi^2$ | 0 | 0 |
| | $a_2$ | 1/6 | 0 | 0 | 1/6 | 0 | 0 | 3/20 | 0 | 0 | $\frac{2(4-\pi)}{\pi^2}$ | 0 | 0 |
| $\boldsymbol{D}^*$ | $a_1$ | 2/5 | 3/5 | 1/10 | 11/35 | 3/7 | 8/35 | 11/42 | 1/3 | 1/3 | 1/2 | $\frac{3(64-2\pi-8\pi^2+\pi^3)}{2\pi^3}$ | $\frac{(-6+\pi^2)}{2\pi^2}$ |
| | $a_2$ | -1 | -3/2 | -1/2 | -4/5 | -6/5 | -4/5 | -7/10 | -1 | -1 | $\frac{12(32-8\pi-\pi^2)}{\pi^3}$ | $\frac{(192+24\pi^2-4\pi^3)}{2\pi^3}$ | $\frac{12(8-\pi^2)}{\pi^3}$ |
| $\boldsymbol{I_0^*}$ | $b_1$ | -1/2 | -1/2 | -1/2 | -1/3 | -1/3 | -2/3 | -1/4 | -1/4 | -3/4 | $-2/\pi$ | $\frac{2-\pi}{\pi}$ | $-2/\pi$ |
| $\boldsymbol{I_1^*}$ | $b_1$ | 1/12 | 0 | 0 | 1/12 | 0 | 0 | 3/40 | 0 | 0 | $\frac{4-\pi}{\pi^2}$ | 0 | 0 |
| $\boldsymbol{I_2^*}$ | $b_1$ | -1/2 | -3/4 | -1/4 | -2/5 | -3/5 | -2/5 | -7/20 | -1/2 | -1/2 | $\frac{6(32-8\pi-\pi^2)}{\pi^3}$ | $\frac{(-48+6\pi^2-\pi^3)}{\pi^3}$ | $\frac{6(8-\pi^2)}{\pi^3}$ |

For an FGP plate with a constant thickness of $h$, total area of $A_{tot}$, total mass of the plate, $M$, is a function of the dimensionless moment of inertia $I_0^*$ and can be calculated as follows:

$$M = \int_{A_{tot}} \int_{-h/2}^{+h/2} \rho(z) dz = A_{tot}.I_0 = A_{tot}.\bar{\rho}.h.I_0^* \qquad (11)$$

Hence, the comparison of masses of FGP plates is similar to comparison of $I_0^*$ in Eq. (10d). Optimizing the performance of an FGP plate as a structural component by use of an identical amount of mass is one of the engineering design goals. In this case, it is necessary to provide a unique value of mass for different porosity distributions which results in selecting different values of maximum porosity parameter $p_m$ for different types of porosity distributions. It should be taken into account that the upper limit for the maximum porosity parameter is one ($p_m < 1$) and the equality of mass should not result in more than one for it. In order to ensure that this condition is met, the plate with the largest dimensionless moment of inertia, $I_0^*$, i.e. S3 or P3 is chosen as the reference. Afterwards the maximum porosity parameter of $p_m$ can be determined by equating $I_0^*$ for all types of porosity distributions with the reference one. Using Table 2, the relationship between the maximum porosity parameter of $p_m$ and the one for the reference ($\overline{p}_m$) can be found as $p_m = \delta \bar{p}_m$ where $\delta$ values are listed in Table 3 for all types of porosity distributions. As an example, if the maximum porosity parameter for P3 be $p_m = 0.8$, the maximum porosity parameters for P1and P2 need to be respectively set to $p_m = 0.4$ ($\delta = 0.5$) and $p_m = 0.6$ ($\delta = 0.75$). Then, for an optimization process, one can be sure that all of them have the same mass.

**Table 3:** Values of $\delta$ for equivalency of mass for different types of porosity distributions.

| | Linear | | | Parabolic | | | Cubic | | | Cosine | | |
|---|---|---|---|---|---|---|---|---|---|---|---|---|
| | P1 | S1 | D1 | P2 | S2 | D2 | P3 | S3 | D3 | PC | SC | DC |
| $\delta$ | 1/2 | 1/2 | 1/2 | 3/4 | 3/4 | 3/8 | 1 | 1 | 1/3 | $\frac{\pi}{8}$ | $\frac{\pi}{4(\pi-2)}$ | $\frac{\pi}{8}$ |

## 3. Results and discussion

Eqs. (8) and (9), together with the coefficients set out in Table 2, can be used to calculate the dimensionless stiffnesses and the dimensionless moments of inertia for various types of porosity distributions for a certain value of the maximum porosity parameter, $p_m$. Afterwards knowing the thickness of the plate, $h$, and properties of the bulk material, i.e. $\bar{\rho}$, $\bar{E}$ and $\bar{\nu}$, the stiffnesses and the moments of inertia can be evaluated by implementing Eq. (7). Considering the geometry, the loading and boundary conditions and material properties of the FGP plate, one needs to look for an analytical analysis among a huge number of solutions based on the well-known theories of plates widely presented in the literature. Then, the stiffnesses and the moments of inertia can be substituted in the closed-form solution. For the case of lacking the exact closed-form solution, the calculated coefficients may be imported to the governing equations of motion to implement a suitable numerical or analytical approach. In order to find the appropriate exact solution for FGP plates, it is recommended to put under consideration two essential questions: Is the FGP plate thin enough to be analyzed using CPT? Are the material properties of the FGP plate is symmetric with respect to the midplane (i.e. S- and D-types)?

The case of thin FGP plates of symmetric properties is an especial situation where the governing equation of plate is the simplest one and the only difference between plates is reflected in the bending stiffness, $D$, and the moment of inertia $I_0$. Therefore, the closed-form solutions for the homogeneous counterpart based on CPT, which are extensively presented in the classical references (For example see [1-3]), may be found and the solution for the FGP plate is simply obtained by multiplying a 'porosity correction factor' which is related to the dimensionless coefficients $D^*$ and $I_0^*$. Bending and free vibration of thin FGP plates of symmetric properties as two examples are presented in Appendix A. Influence of type of porosity distribution on the response of FGP plate for different boundary conditions, loading, and the geometry of plate is discussed in detail. In contrast, for both thick FGP plates and thin FGP plates of non-symmetric properties (P-type porosity distribution), there is a system of equations instead of a single equation. However, one can find many exact closed-form solutions for thick plates, mostly based on FSDT. It is noted that the solutions for non-homogenous plates such as FGMs and nanocomposite plates as well as orthotropic plates may be applicable. Appendix B and Appendix C respectively provide two instances for implementing closed-form solutions to investigate the bending of FGP thin plates of non-symmetric properties and FGP thick plates, respectively.

For validating the results presented in Appendixes, the outcomes from analytical solutions have been compared to the ones from FEM commercial software, COMSOL Multiphysics©. In order to model the FGP plates in the software, shell physic has been used with at least 20 elements in the thickness direction while the quadratic type of meshes has been set. The results are reported for $p_m = 0.99$ to consider the maximum influence of porosity on the response of the FGP plates. It should be noted that the percent of the difference between the results obtained by analytical solutions and the ones from FEM is calculated as:

$$\%Error = \left|\frac{Plate\ Theory - FEM}{FEM}\right| \times 100 \quad (13)$$

Finally, to make clear the proposed approach, the flowchart for the steps needs to be followed is depicted in Fig. 3.

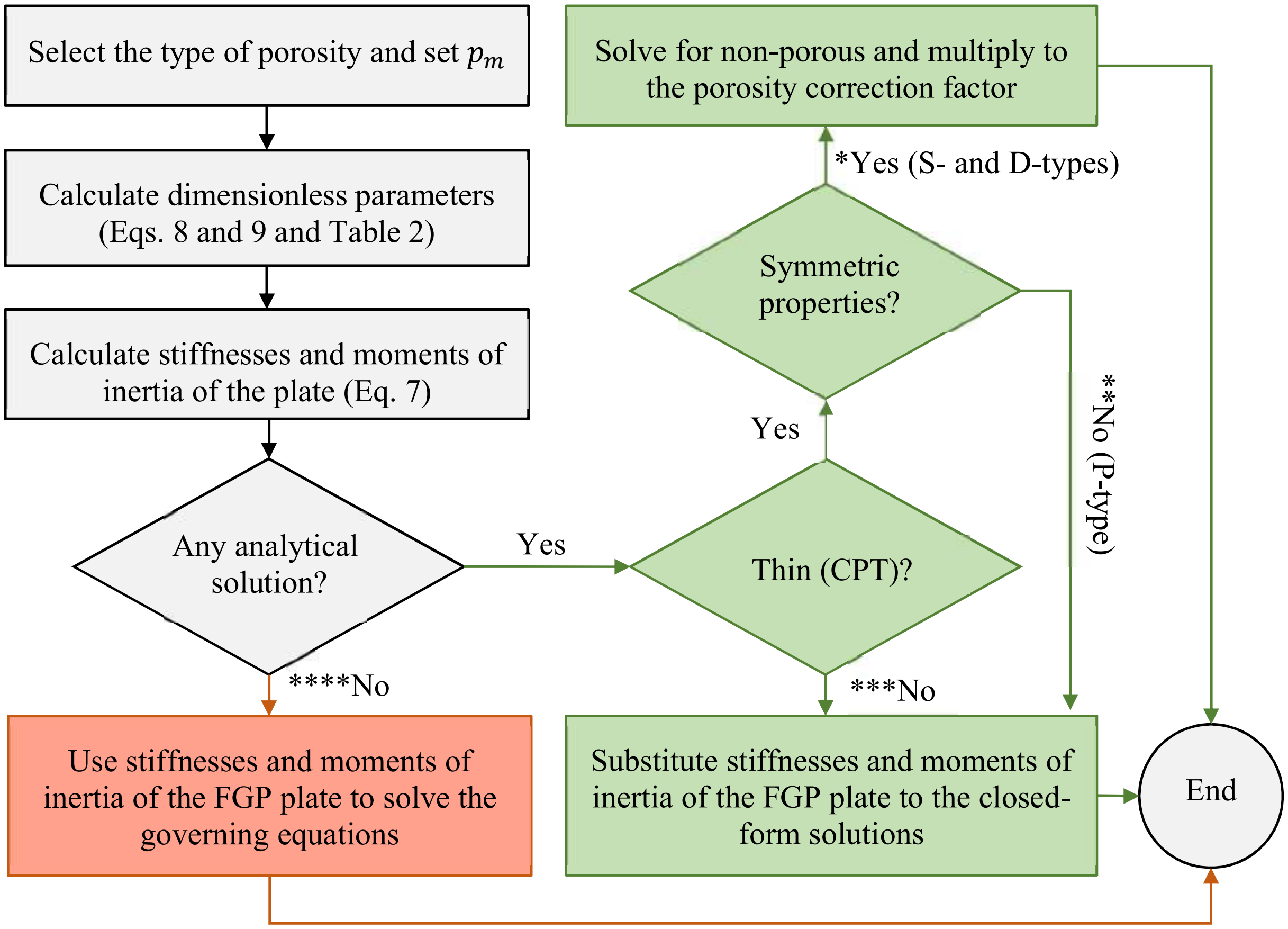

* No need to resolve. See Appendix A for some examples.
** No need to resolve. See Appendix B for an example.
*** No need to resolve. See Appendix C for an example.
**** Needs to solve the governing equations of FGP plates.

**Fig 3.** Proposed flowchart for analyzing mechanical behavior of FGP plates.

## 4. Conclusion

This paper presents a guideline for implementing the available analytical closed-form solutions in the literature of plate theories for analyzing the mechanical responses of FGP plates. 12 different types of functionality covering a wide range of porosity distributions along the thickness are considered. For every distribution, explicit formulations for calculating stretching, bending-stretching, and bending stiffnesses as well as moments of inertia are presented. Then, it is shown that how these coefficients can be substituted into the exact solutions to study the effect of porosity distribution on the mechanical behavior of FGP plates in details through four examples as follows:

- Bending of square and circular thin FGP plates of symmetric distribution of porosity considering combination of clamped and simply supported edges under uniform or hydrostatic transverse loading using CPT.
- Free vibration of triangular, parallelogram, annular, and annular sectorial thin FGP plates of symmetric distribution of porosity for either clamped or simply supported edges using CPT.
- Bending of thin simply supported square FGP plates of non-symmetric distribution of porosity using CPT.
- Bending of thick simply supported square FGP plates of symmetric and non-symmetric distribution of porosity based on FSDT.

It is approved that the mechanical response of the first two examples i.e. the case of thin FGP plates with symmetric properties with respect to midplane can be simply obtained by multiplying a 'porosity correction factor' to the closed-form solution of the homogeneous counterpart. All the presented results are compared to FEM simulations using COMSOL Multiphysics© software to validate the accuracy of the predicted results.

## Data Availability

The raw/processed data required to reproduce these findings cannot be shared at this time due to technical or time limitations.

## Appendix A. Bending and free vibration of symmetric thin FGP plates

In this appendix, the bending and free vibration responses of thin FGP plates of symmetric porosity distribution (D- and S-types) are presented. Due to symmetry the coefficients of $B$ and $I_1$ are zero. Based on CPT the out-of-plane governing equation of motion is decoupled from in-plane motion and can be analyzed separately as:

$$D\nabla^4 w - q = I_0 \ddot{w} \qquad \text{(A.1)}$$

where $q$ and $w$ are the distributed transverse loading and the out of plane displacement, respectively. $\nabla^4$ is the Laplace operator and $(\dot{\ })$ represent derivative with respect to time.

### A.1. Bending of Symmetric FGP Plates

The governing equation for the symmetric FGP plates under static bending is:

$$\nabla^4 w = \frac{q}{D}, \qquad (w \propto \frac{1}{D}) \qquad \text{(A.2)}$$

It can be concluded that the deflection of plate is inversely proportional to the bending stiffness. The deflection porosity correction factor, $\gamma_w$, is introduced as the ratio of the deflection of the FGP plate, $w$, to the deflection of its non-porous counterpart, $\bar{w}$. Considering Eqs. (A2) and (7), the dimensionless deflection parameter of $\gamma_w$ is simply evaluated by inversing the dimensionless bending stiffness $D^*$ as follows:

$$\gamma_w = \frac{w}{\bar{w}} = \frac{\bar{D}}{D} = \frac{1}{D^*} \qquad \text{(A.3)}$$

In the other word the deflection of thin FGP plate of symmetric porosity distribution, $w$, can be obtained by multiplying the deflection of its homogeneous counterpart, $\bar{w}$, to the deflection porosity correction factor, $\gamma_w$.

In order to validate Eq. (A3), in Figs. A1 to A3 the outcomes from FEM have been compared to the corresponding values of $1/D^*$ for symmetric distributions of porosity along the thickness. It is noted that the values on the top of each bar indicate the error in percentage based on Eq. (13). Fig. A1 illustrates the deflection porosity correction factor of an FGP circular plate with a diameter of $a$, thickness of $h$ under a uniform transverse loading for either clamped or simply supported boundary conditions. Two thickness to side-length ratios, $h/a$ equals to 0.01 and 0.1 is considered to also estimate the amount of error of CPT for thicker plates. It is obvious that the deflection porosity correction factor is always greater than one which means that the deflection of the FGP plate is greater than its non-porous counterpart. The maximum deflection porosity correction factor relates to S1 while the minimum one is to D1. The deflection of the porous plate with maximum porosity of $p_m = 0.99$ and $h/a$=0.01 is 9.7 times bigger than the non-porous one for S1 case while for the D1 case, it is just 1.66. This is because the higher values of porosity far from the mid-plane lead to the less bending stiffness, so the deflection of the FGP plate increases. In addition, the effect of porosity on the deflection for both clamped and simply supported boundary conditions is the same. Investigating the percentage of error on the top of each bar, it can be deduced that for the circular plates with $h/a$=0.01, the predicted values of $1/D^*$ from the CPT are in very good agreement with the deflection porosity correction factor from FEM. However, as expected that for the thicker plate of $h/a$=0.1, the error increases which is bigger for the clamped boundary condition than simply supported one. Besides, for thick plates, the CPT overestimates the deflection outcomes for S-type while it underestimates the deflections for D-types. It can be concluded that using the CPT for the thick plate can only give an initial inaccurate prediction while for the precise results, the FSDT should be taken into account.

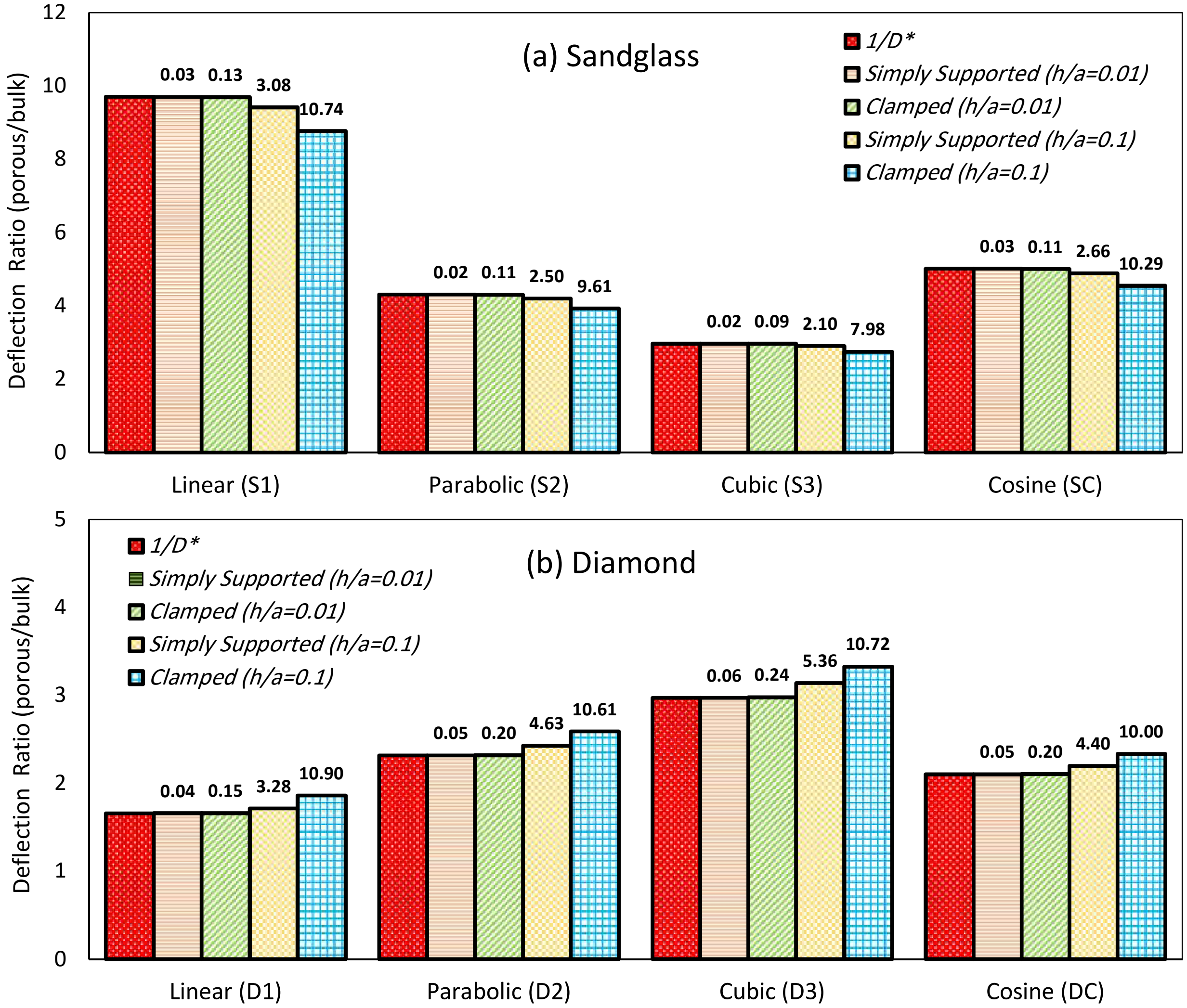

**Fig A1**. Deflection porosity correction factor for simply supported and clamped circular plates under the transverse uniform distributed loading. a) Sandglass, b) Diamond. The values on the top of bars demonstrate the percentage of errors via Eq. (13).

The effect of porosity distribution on the deflection porosity correction factor for FGP square plates with the length of $a$ and the thickness of $h$ is shown in Fig. A2. The behavior of the square plate is very similar to the circular one and the CPT can accurately predict the deflection of FGP square thin plates. In addition to the symmetric boundary condition with four simply supported edges (SS-SS-SS-SS), the boundary condition with three simply supported edges and one clamped edge (SS-SS-SS-C) has been taken into account. The results show that the combined boundary condition for thick plate results in a larger error than in the symmetric one.

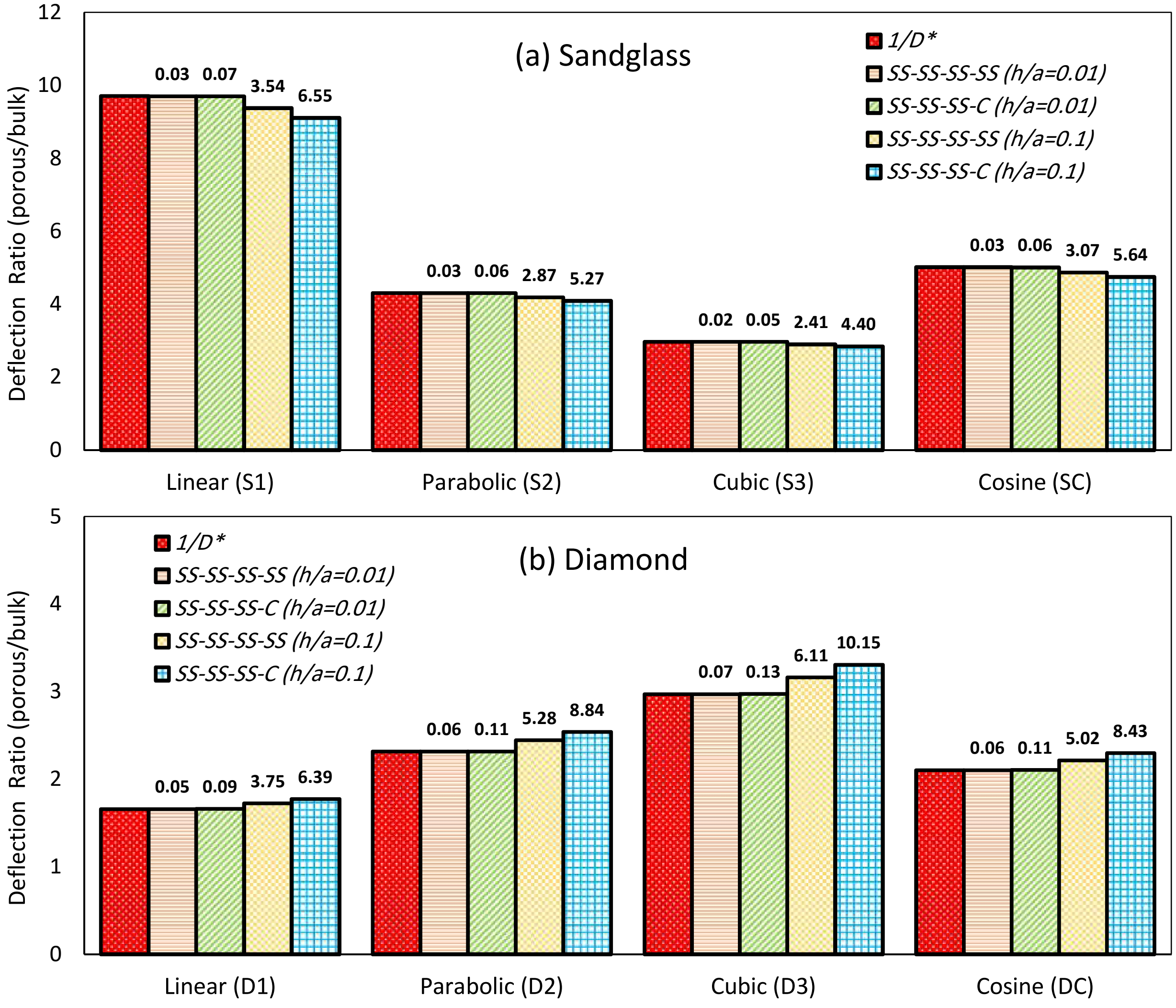

**Fig A2**. Deflection porosity correction factor for SS-SS-SS-SS and SS-SS-SS-C square plates under the transverse uniform distributed loading. a) Sandglass, b) Diamond. The values on the top of bars demonstrate the percentage of errors via Eq. (13).

To check the validity of Eq. (A3) for non-uniform loading, hydrostatic loading is applied to the FGP square plates. Fig. A3 indicates that the CPT presented in Eq. (A3) is also in good agreement with FEM results for thin plates while for thick FGP plates overestimates and underestimates dimensionless deflection for S- and D-types of porosity distributions, respectively.

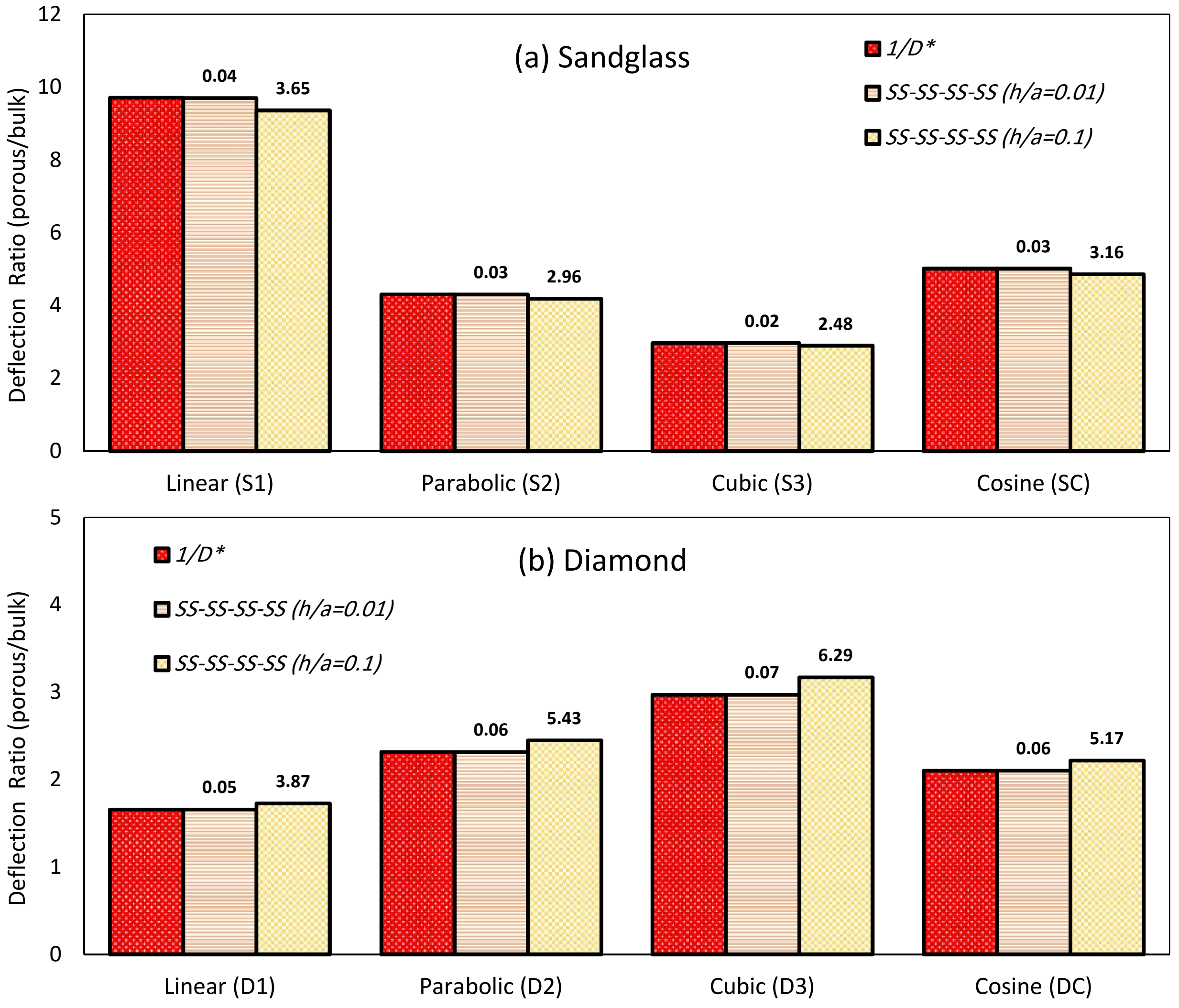

**Fig. A3**. Deflection porosity correction factor for SS-SS-SS-SS square plates under the transverse hydrostatic distributed loading. a) Sandglass, b) Diamond. The values on the top of bars demonstrate the percentage of errors via Eq. (13).

### *A.2. Free Vibration of Symmetric FGP Plates Based on CPT*

Eq. (A1) can be rewritten for the free vibration of FGP plates in transverse direction by assuming the harmonic motion as:

$$\nabla^4 w = -\Omega^2 \frac{I_0}{D}, \qquad (\Omega \propto \sqrt{D/I_0}) \tag{A.4}$$

From Eq. (A4) one can see that for symmetric distributions of porosity, the natural frequency, $\Omega$, is proportional to the square root of the ratio of bending stiffness to the first moment of inertia. To study the free vibration of the FGP plate based on the CPT, the frequency porosity correction factor, $\gamma_\Omega$, has been introduced as the ratio of the natural frequency of the FGP plate to natural frequency of its non-porous counterpart, $\bar{\Omega}$. Considering Eq. (A4), the frequency porosity correction factor is related to the dimensionless bending stiffness and the dimensionless first moment of inertia as:

$$\gamma_{\Omega} = \frac{\Omega}{\bar{\Omega}} = \sqrt{D\bar{I}_0/\bar{D}I_0} = \sqrt{D^*/I_0^*} \tag{A.5}$$

It can be concluded that the natural frequencies of thin FGP plate of symmetric porosity distribution, $\Omega$, are obtained by multiplying the frequencies of its non-porous counterpart, $\bar{\Omega}$, to the frequency porosity correction factor, $\gamma_{\Omega}$.

To validate the accuracy of Eq. (A5), at following the frequency porosity correction factor of the FGP plate by means of FEM has been presented for different geometrical shapes, considering either simply supported or clamped boundary conditions for thin and thick plates of $h/a$ equal to 0.01 and 0.1 and have been compared to the corresponding parameter of $\sqrt{D^*/I_0^*}$. The values on the top of the bars show the percentage errors according to Eq. (13).

The frequency porosity correction factor of the FGP plates for the equilateral triangular, parallelogram, annular, and annular sectorial shapes are respectively demonstrated in Figs. A4 to A7. One can see that the variation of the frequency porosity correction factor with respect to porosity distribution is almost the same for all the plate shapes and the results obtained for the thin FGP plates using Eq. (A5) show excellent accuracy compare to the FEM results. It is seen that the frequency porosity correction factor for S-type porosity distribution is always less than one. It means that purposefully implementation of the S-type distribution can decrease the natural frequency with respect to its non-porous counterpart. The decrease in frequency is related to S1 (55%), while the minimum is shown for S2 (41%). It should be noted that these values are for the maximum porosity parameter of $p_m = 0.99$. The amount of this reduction in frequency can be tuned by changing the maximum porosity parameter, $p_m$, in order to avoid resonance. The reason for reducing the frequency is that the porosity decreases both the bending stiffness and the first moment of inertia while for S-type distribution, reduction of stiffness is more than the moment of inertia. On the other side, for D-type distribution, the frequency porosity correction factor is slightly greater than one which means that these types of FGP plates have a greater natural frequency than the non-porous one. This is because the porosity leads to more reduction in the first moment of inertia than the bending stiffness. For a plate with $h/a$=0.01 and maximum porosity parameter of $p_m = 0.99$, the maximum of increase in frequency belongs to D3 (14%) and the minimum is for the D1 (9%). It can be seen that for all eight types of porosity distributions, the frequency porosity correction factor is the same for both clamped and simply supported boundary conditions. Analyzing the errors on the top of the bars shows that the CPT results for thin plates of $h/a$=0.01 have great accuracy compare to the FEM ones. Meanwhile, as it was expected, for $h/a$=0.1, the CPT results are not correlated with FEM ones with high accuracy and the clamped boundary condition have less precise outcomes than the simply supported ones. Unlike the results for deflection, the CPT results for thick plate underestimates the natural frequency results for S-type distribution while overestimates the natural frequency outcomes for D-type one.

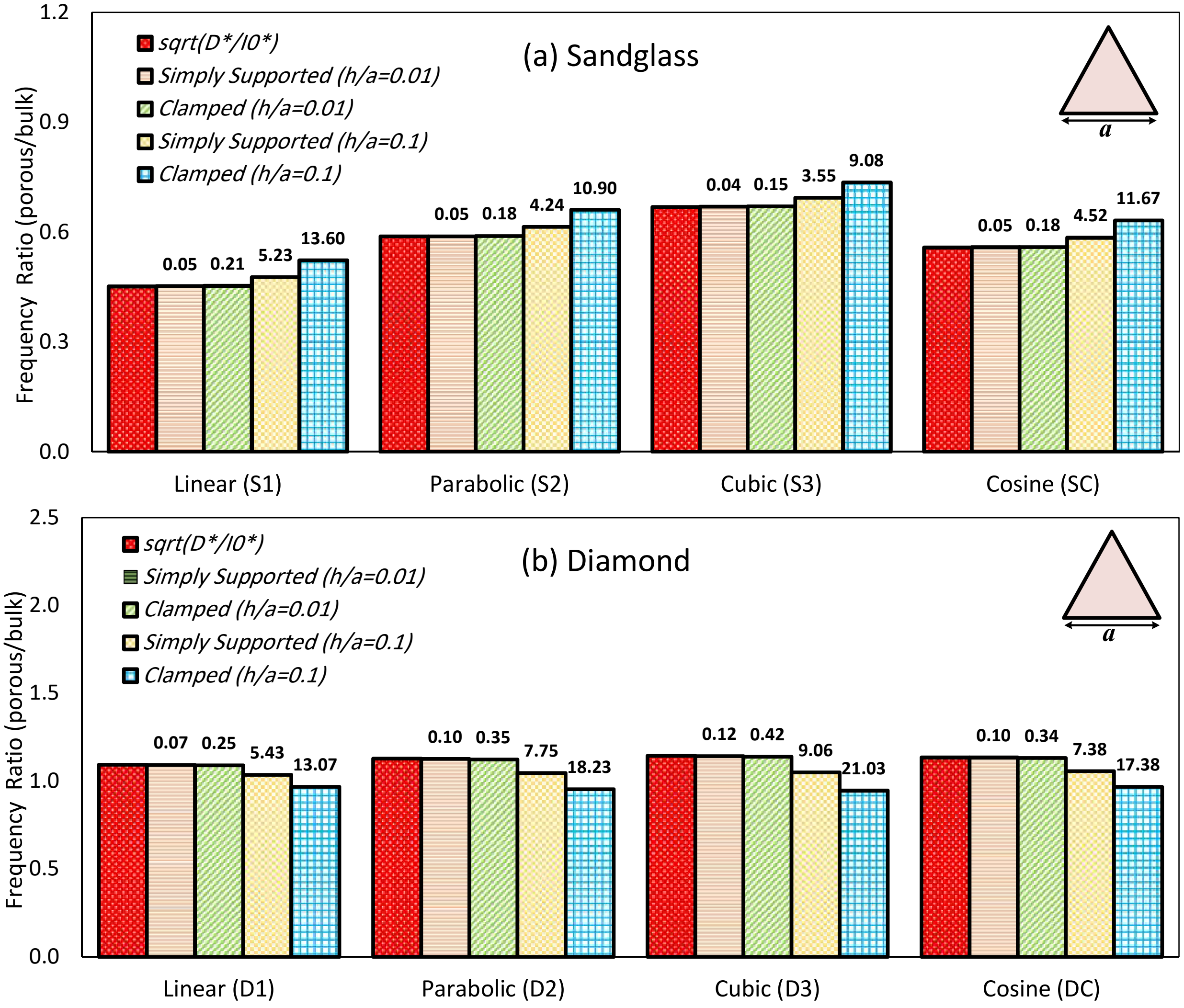

**Fig A4**. Frequency porosity correction factor for simply supported and clamped equilateral triangular plates. a) Sandglass, b) Diamond. The values on the top of bars demonstrate the percentage of errors via Eq. (13). Results are presented based on fundamental natural frequency.

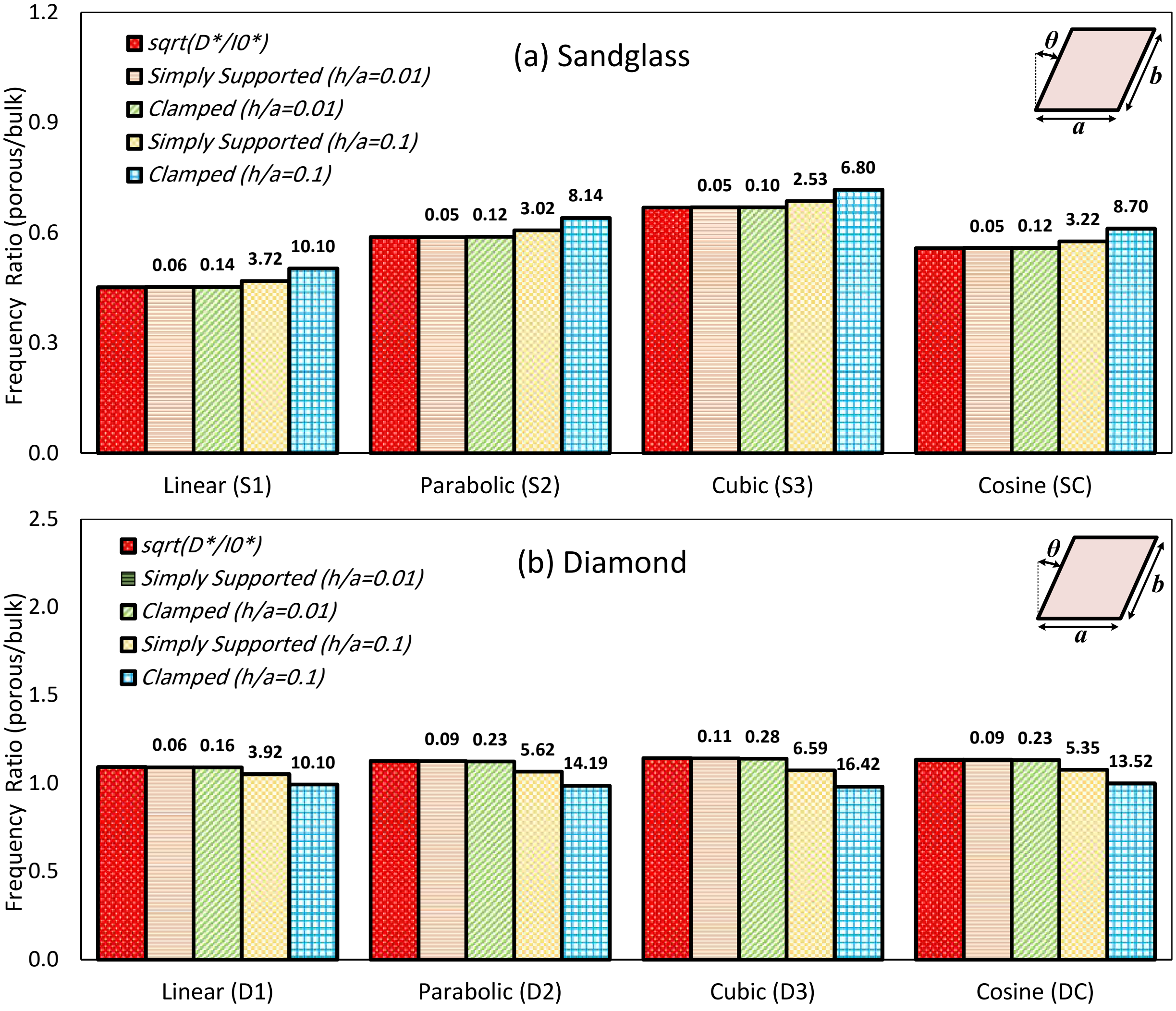

**Fig A5**. Frequency porosity correction factor for simply supported and clamped parallelogram plates, $a/b = 1$, $\theta = 45°$. a) Sandglass, b) Diamond. The values on the top of bars demonstrate the percentage of errors via Eq. (13). Results are presented based on fundamental natural frequency.

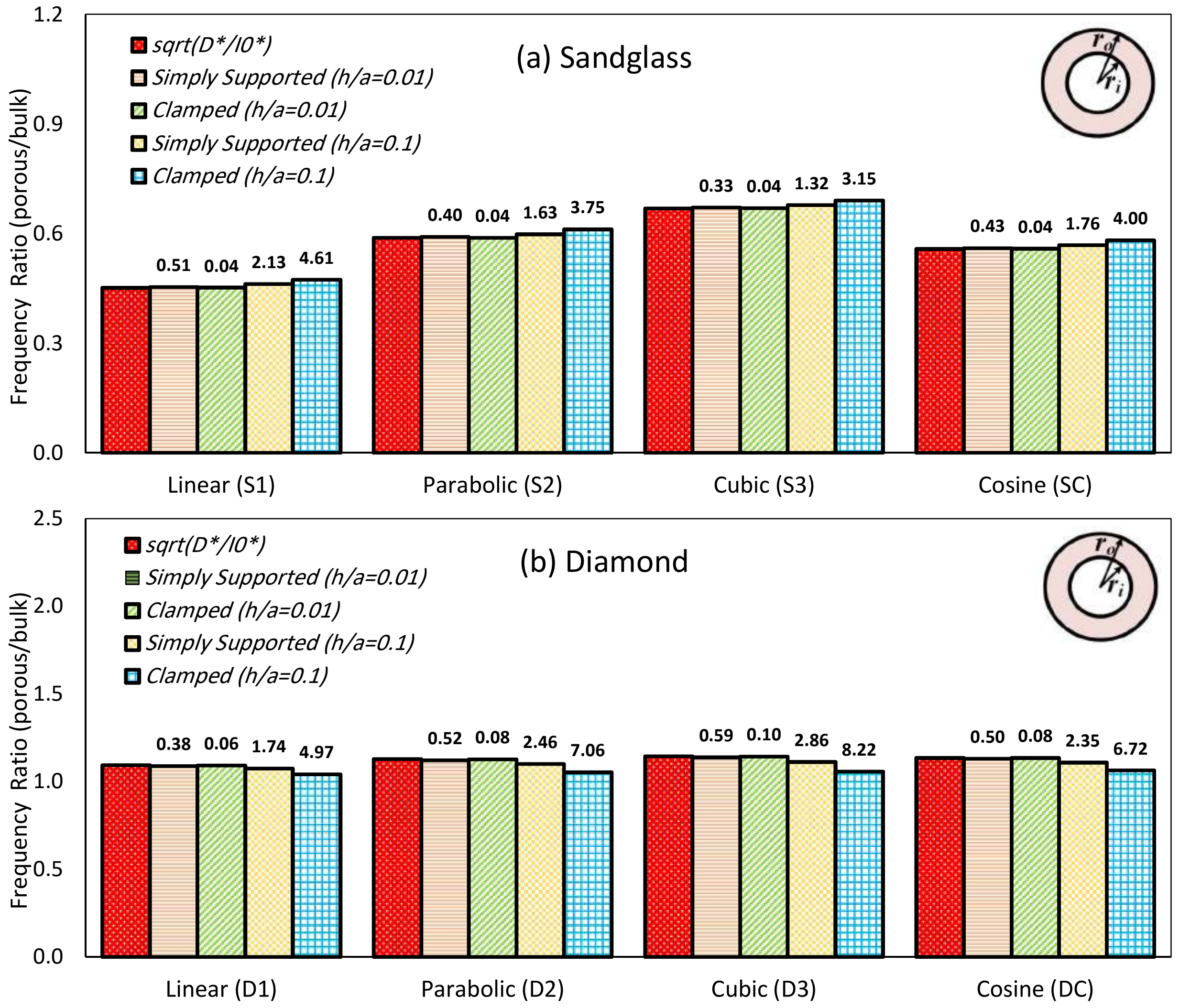

**Fig A6**. Frequency porosity correction factor for simply supported and clamped annular plates, $r_i/r_o = 2,\ a = r_0 - r_i$. a) Sandglass, b) Diamond. The values on the top of bars demonstrate the percentage of errors via Eq. (13). Results are presented based on fundamental natural frequency.

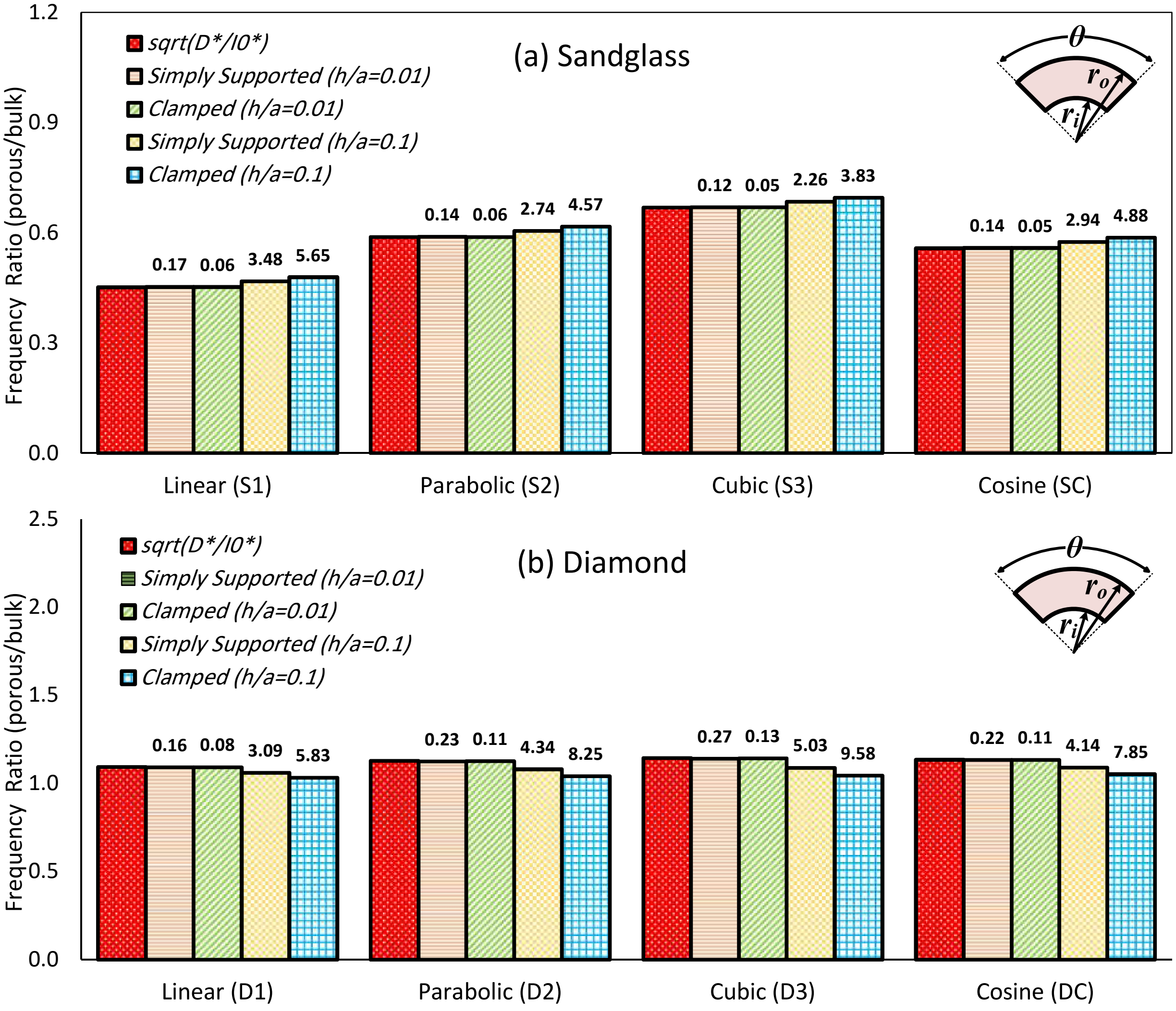

**Fig A7**. Frequency porosity correction factor for simply supported and clamped annular sectorial plates, $r_i/r_o = 2$, $a = r_0 - r_i$, $\theta = 90°$. a) Sandglass, b) Diamond. The values on the top of bars demonstrate the percentage of errors via Eq. (13). Results are presented based on fundamental natural frequency.

## Appendix B. Bending of non-symmetric thin FGP plates

In this appendix, the bending of thin FGP plates of non-symmetric properties with respect to the midplane, P-type, has been presented. The governing equations of motions for the isotropic nonhomogeneous plate based on the CPT are as follows [6]:

$$A\left[u_{,xx} + \bar{v}v_{,xy} + 0.5(1-\bar{v})\left(u_{,yy} + v_{,xy}\right)\right] = 0 \tag{B.1}$$

$$A\left[\bar{v}u_{,xy} + v_{,yy} + 0.5(1-\bar{v})\left(u_{,xy} + v_{,xx}\right)\right] = 0 \tag{B.2}$$

$$B\left[u_{,xxx} + v_{,yyy} + \bar{v}\left(u_{,xyy} + v_{,xxy}\right) + (1-\bar{v})\left(u_{,xyy} + v_{,xxy}\right)\right] + D\nabla^4 w - q = 0 \tag{B.3}$$

where $u$ and $v$ are the displacement of the mid-plane along the in-plane directions, $x$ and $y$, and $q$ and $w$ are the distributed transverse loading and the out of plane displacement, respectively. The implicit derivative in term of in-plane coordinates is denoted by $(\ )_{,x}$ and $(\ )_{,x}$. Based on the CPT, perpendicular planes to the mid-plane are not deformed after loading and remain perpendicular to the mid-plane which means that $\psi_x$ and $\psi_y$ which are the rotations around the in-plane axes are equal to $\psi_x = -w_{0,x}$ and $\psi_y = -w_{0,y}$. Therefore, $\psi_x$ and $\psi_y$ do not appear as independent variables in Eqs. (B1) to (B3) which is a system of three differential equations with three unknown variables.

For antisymmetric material properties respect to the mid-plane, P-type, the coefficients of $B$ and $I_1$ are not zero so that there is a coupling between the in-plane displacement (*u* and *v*) and out-of-plane one (*w*). Thus, analyzing the response of inhomogeneous plate with antisymmetric material properties require to solve the system of Eqs. (B1) to (B3) in order to determine the displacement field simultaneously. There are many analytical and numerical solutions for solving this system of equations for different loading and boundary conditions.

In the following, the bending behavior of the FGP square plates of the side length of *a* having antisymmetric porosity distribution, P-type, and simply supported boundary conditions under uniform transversal load $q$ has been examined based on the CPT analytical solution in the literature. The Navier solution for a rectangular plate of dimensions of *a* and *b* have been reported in Ref. [6]. The deflection of the plate can be calculated based on the followed series:

$$w = \sum_{n=1}^{\infty}\sum_{m=1}^{\infty} W_{mn}\ sin(\alpha x)\ sin(\beta y) \tag{B.4}$$

In which $\alpha = m\pi x/a$ and $\beta = n\pi/b$ and $W_{mn}$ can be found as follow:

$$\begin{aligned} W_{mn} &= \frac{Q_{mn}}{a_{mn}}, Q_{mn} = \frac{16q}{mn\pi^2} \\ a_{mn} &= c_{33} + c_{13}\frac{a_1}{a_0} + c_{23}\frac{a_2}{a_0} \\ a_0 &= c_{11}c_{22} - c_{12}^2, a_1 = c_{12}c_{23} - c_{13}c_{22}, a_2 = c_{13}c_{12} - c_{11}c_{23} \end{aligned} \tag{B.5}$$

The $c_{ij}$ are related to the plate stiffnesses as follow:

$$\begin{aligned} c_{11} &= (\alpha^2 + 0.5(1-\bar{\nu})\beta^2)A \\ c_{12} &= 0.5(1+\bar{\nu})\alpha\beta A \\ c_{13} &= -(\alpha^3 + \alpha\beta^2)B \\ c_{22} &= (\beta^2 + 0.5(1-\bar{\nu})\alpha^2)A \\ c_{23} &= -(\beta^3 + \alpha^2\beta)B \\ c_{33} &= (\alpha^2 + \beta^2)^2 D \end{aligned} \tag{B.6}$$

The dimensionless deflection is defined as the ratio of porous to non-porous plate. This value for an FGP square plate of the side length of *a*, thickness of *h* with *h*/*a* equal to 0.01 and 0.1 with antisymmetric porosity distribution is presented in Fig. B1. The maximum deflection porosity correction factor is related to PC type while the minimum one is for P3. For the maximum porosity parameter of $p_m = 0.99$ and $h/a$=0.01, the increase in deflection for PC-type is 14.96 times greater than the non-porous one while for P3-type, this increase is 2.57 times. By analyzing the errors on the top of the bars, same as the symmetric cases, the CPT can predict the results for $h/a$=0.01 with high accuracy compared to the FEM results. It is also expected that the results of CPT for thick plates ($h/a$=0.1) present larger errors overestimating the deflection for the P-type. It is seen that the dimensionless parameter, $1/D^*$ is just valid for the symmetric porosity distribution, whereas the dimensionless deflection for the P-type is completely far from this parameter.

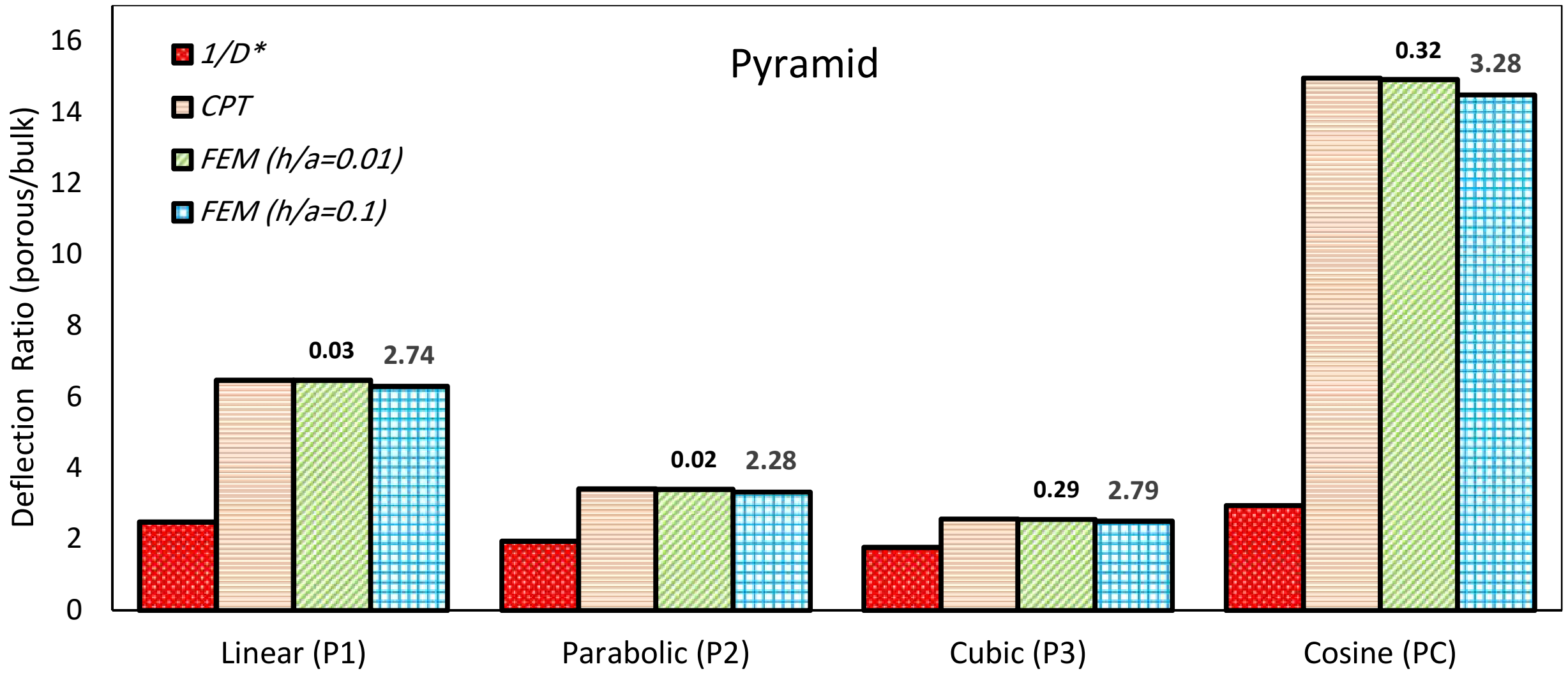

**Fig B1**. Dimensionless deflection for simply supported square plates with Pyramid porosity distributions. The values on the top of bars demonstrate the percentage of errors via Eq. (13).

## Appendix C. Bending of thick FGP plates

In this appendix by implementing the FSDT, the behavior of the thick FGP plates has been examined based on analytical solutions in the literature. In this case, in addition to the three mid-plane displacements $u$, $v$ and $w$, two independent variables of $\psi_x$ and $\psi_y$ are considered for rotations around in-plane axes in the displacement field. The governing equations of motions for FSDT plates includes a system of five differential equations as [4]:

$$A\left[u_{,xx} + \bar{\nu} v_{,yx} + 0.5(1-\bar{\nu})\left(u_{,yy} + v_{,xy}\right)\right] + B\left[\psi_{x,xx} + \bar{\nu}\,\psi_{y,yx} + 0.5(1-\bar{\nu})(\psi_{x,yy} + \psi_{y,yx})\right] = 0 \tag{C.1}$$

$$A\left[\bar{\nu} u_{,xy} + v_{,yy} + 0.5(1-\bar{\nu})\left(u_{,yx} + v_{,xx}\right)\right] + B\left[\bar{\nu}\psi_{x,xy} + \psi_{y,yy} + 0.5(1-\bar{\nu})(\psi_{x,xy} + \psi_{y,xx})\right] = 0 \tag{C.2}$$

$$0.5k_s(1-\bar{\nu})A\left[w_{,xx} + \psi_{x,x} + w_{,yy} + \psi_{y,y}\right] - q = 0 \tag{C.3}$$

$$-0.5k_s(1-\bar{\nu})A\left[w_{,x} + \psi_x\right] + B\left[u_{,xx} + \bar{\nu} v_{,yx} + 0.5(1-\bar{\nu})\left(u_{,yy} + v_{,xy}\right)\right] + D\left[\psi_{x,xx} + \bar{\nu}\,\psi_{y,yx} + 0.5(1-\bar{\nu})\left(\psi_{x,yy} + \psi_{y,yx}\right)\right] = 0 \tag{C.4}$$

$$-0.5k_s(1-\bar{v})A\left[w_{,y}+\psi_y\right]+B\left[\bar{v}u_{,xy}+v_{,yy}+0.5(1-\bar{v})\left(u_{,yx}+v_{,xx}\right)\right] + D\left[\bar{v}\psi_{x,xy}+\ \psi_{y,yy}+0.5(1-\bar{v})\left(\psi_{x,xy}+\ \psi_{y,xx}\right)\right]=0 \tag{C.5}$$

where $k_s$ is the shear correction factor of FSDT which is considered to be 5/6. As the FSDT considers the effects of the shear deformation, its outcomes for the thick plate is also accurate and converge to the ones from the three-dimensional elasticity theory. Many references addressed the analytical solution of the system of Eqs. (C1) to (C5) for different loading and boundary condition especially the ones related to FGM made of metal and ceramic in the recent two decades.

Here as an example, the bending behavior of an FGP square plate with the side length of $a$ of simply supported boundary condition under uniform transversal loading $q$ has been investigated. The general Navier solution for the FGP plate having either symmetric or antisymmetric porosity distributions based on the FSDT has been available in the Ref. [6]. The displacement field components of the plate can be assessed by the following series:

$$(u,v,w,\psi_x,\psi_y)=\sum_{n=1}^{\infty}\sum_{m=1}^{\infty}(U_{mn},V_{mn},W_{mn},X_{mn},Y_{mn},)\ sin(\alpha x)\,sin(\beta y) \tag{C.6}$$

where $(U_{mn},V_{mn},W_{mn},X_{mn},Y_{mn},)$ can be determined as follow:

$$\begin{Bmatrix} U_{mn}\\ V_{mn}\\ W_{mn}\\ X_{mn}\\ Y_{mn}\end{Bmatrix}=\begin{bmatrix} k_{11} & k_{12} & 0 & k_{14} & k_{15}\\ k_{12} & k_{22} & 0 & k_{24} & k_{25}\\ 0 & 0 & k_{33} & k_{34} & k_{35}\\ k_{14} & k_{24} & k_{34} & k_{44} & k_{45}\\ k_{15} & k_{25} & k_{35} & k_{45} & k_{55}\end{bmatrix}^{-1}\begin{Bmatrix} 0\\ 0\\ Q_{mn}\\ 0\\ 0\end{Bmatrix} \tag{C.7}$$

$Q_{mn}=\frac{16q}{mn\pi^2}$.

The $k_{ij}$ coefficients are related to the plate stiffnesses and one can obtain them as follow:

$$\begin{aligned}
k_{11}&=(\alpha^2+0.5(1-\bar{v})\beta^2)A\\
k_{12}&=0.5(1+\bar{v})\alpha\beta A\\
k_{14}&=(\alpha^2+0.5(1-\bar{v})\beta^2)B\\
k_{15}&=0.5(1+\bar{v})\alpha\beta B\\
k_{22}&=(\beta^2+0.5(1-\bar{v})\alpha^2)A\\
k_{24}&=k_{15}\\
k_{25}&=(\beta^2+0.5(1-\bar{v})\alpha^2)B\\
k_{33}&=0.5k_s(1-\bar{v})(\alpha^2+\beta^2)A\\
k_{34}&=0.5k_s(1-\bar{v})\alpha A\\
k_{35}&=0.5k_s(1-\bar{v})\beta A\\
k_{44}&=(\alpha^2+0.5(1-\bar{v})\beta^2)D+0.5k_s(1-\bar{v})A
\end{aligned} \tag{C.8}$$

$$k_{45} = 0.5(1+\bar{\nu})\alpha\beta D$$
$$k_{55} = (\beta^2 + 0.5(1-\bar{\nu})\alpha^2)D + 0.5k_s(1-\bar{\nu})A$$

For S- and D-types with symmetric porosity distribution the bending-stretching stiffness $B$ is equal to zero and the value of $W_{mn}$ can be simply evaluated as follows rather than Eq. (C7):

$$W_{mn} = \frac{Q_{mn}}{b_{mn}}$$
$$b_{mn} = k_{33} + k_{34}\frac{b_1}{b_0} + k_{35}\frac{b_2}{b_0} \tag{C.9}$$
$$b_0 = k_{44}k_{45} - k_{45}^2, b_1 = k_{45}k_{35} - k_{34}k_{55}, b_2 = k_{34}k_{45} - k_{44}k_{35}$$

Figs. C1 to C3 show the dimensionless deflection, the deflection ratio of porous/non porous plate, in terms of the thickness to side length ratio, $h/a$. The dimensionless deflection has been presented based on the CPT, FSDT, and FEM. It is demonstrated that for all types of porosity distribution, good correlation can be seen between the FSDT and FEM results for all values of *h*/*a* from the thin FGP plates (*h*/*a*=0.01) to the completely thick ones (*h*/*a*=0.2). It is seen that the CPT can predict the deflection for *h*/*a*=0.01 with good accuracy while by increasing the thickness of the plate, its results diverge from the FSDT and FEM ones. Also, it is seen that the CPT overestimate the dimensionless deflection for the S- and P-types while underestimate for the D-type. It should be noticed that the CPT results show errors for *h*/*a*=0.2 that is significant and cannot be neglected. One can observe that the values of the deflection correction factor, $1/D^*$ is same as CPT results for symmetric distributions, Figs. C1 and C2, and can predict the deflection correctly only for thin FGP plates, however, cannot be used even for thin FGP plates with antisymmetric porosity distribution, Fig C3 (See the starting point of graphs, *h*/*a*=0.01).

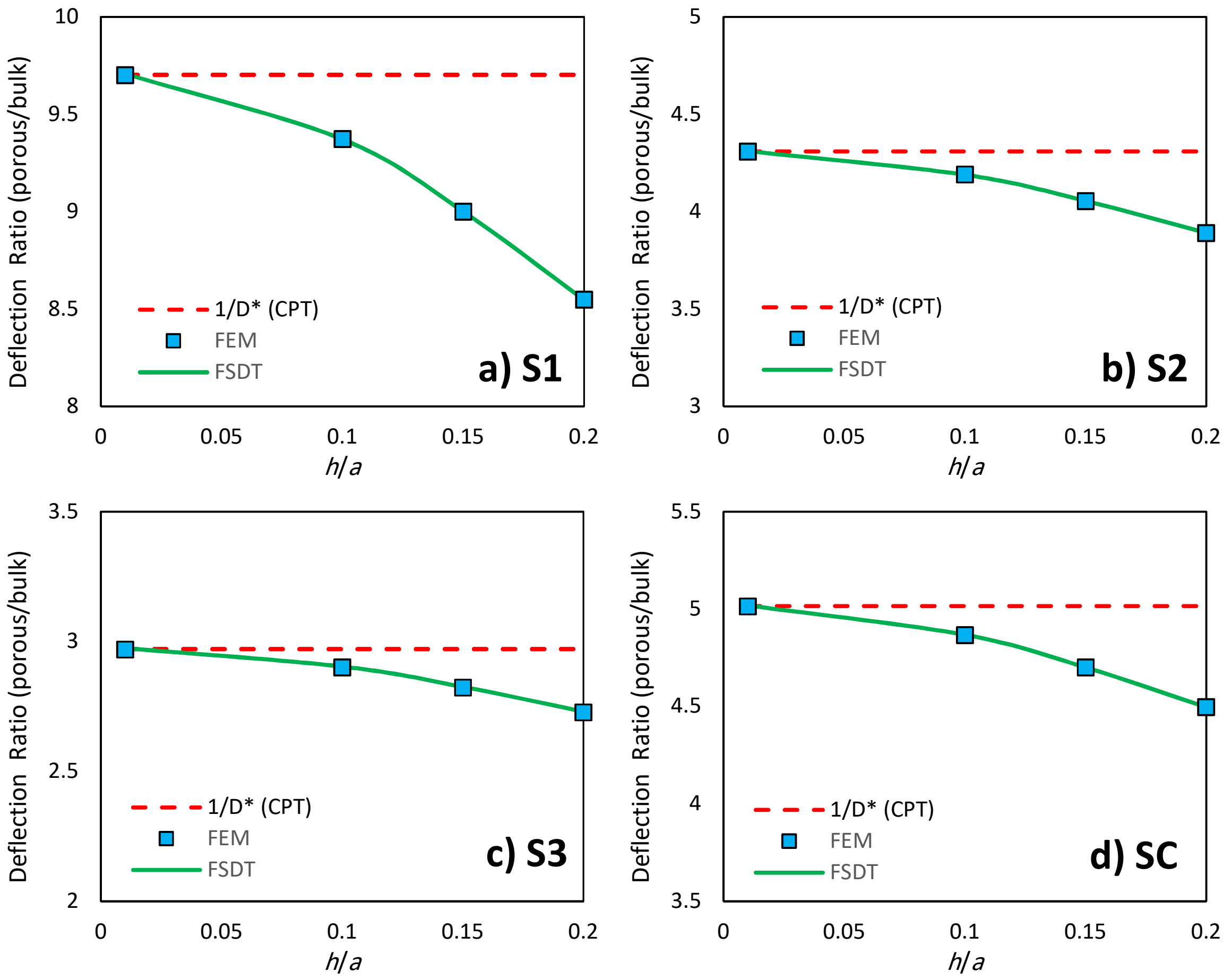

**Fig C1**. Dimensionless deflection for SS-SS-SS-SS square plates versus thickness to side-length ratio under the transverse uniform distributed loading. a) Sandglass Linear S1, b) Sandglass Parabolic S2, c) Sandglass Cubic S3, d) Sandglass Cosine SC.

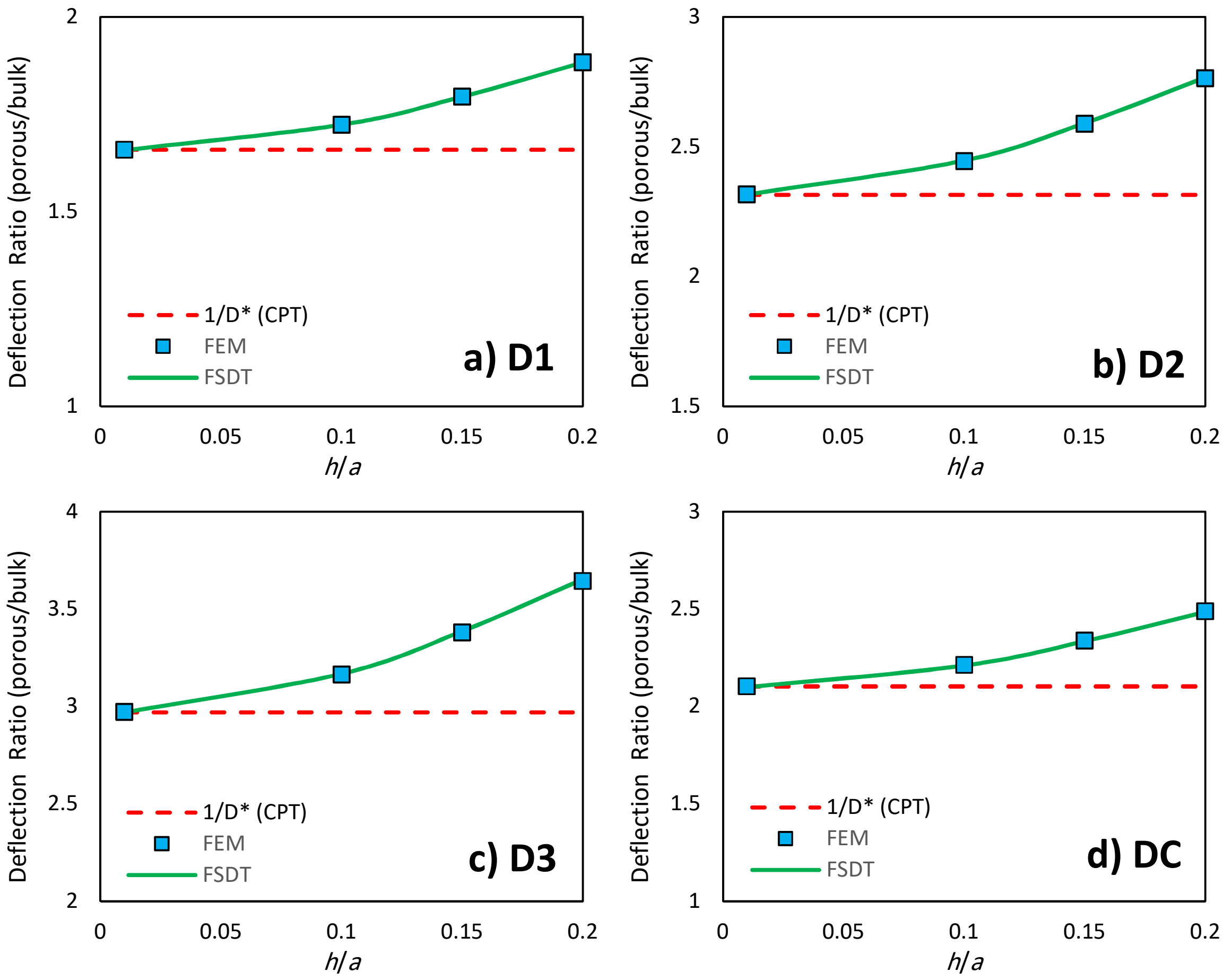

**Fig. C2**. Dimensionless deflection for SS-SS-SS-SS square plates versus thickness to side-length ratio under the transverse uniform distributed loading. a) Diamond Linear D1, b) Diamond Parabolic D2, c) Diamond Cubic D3, d) Diamond Cosine DC.

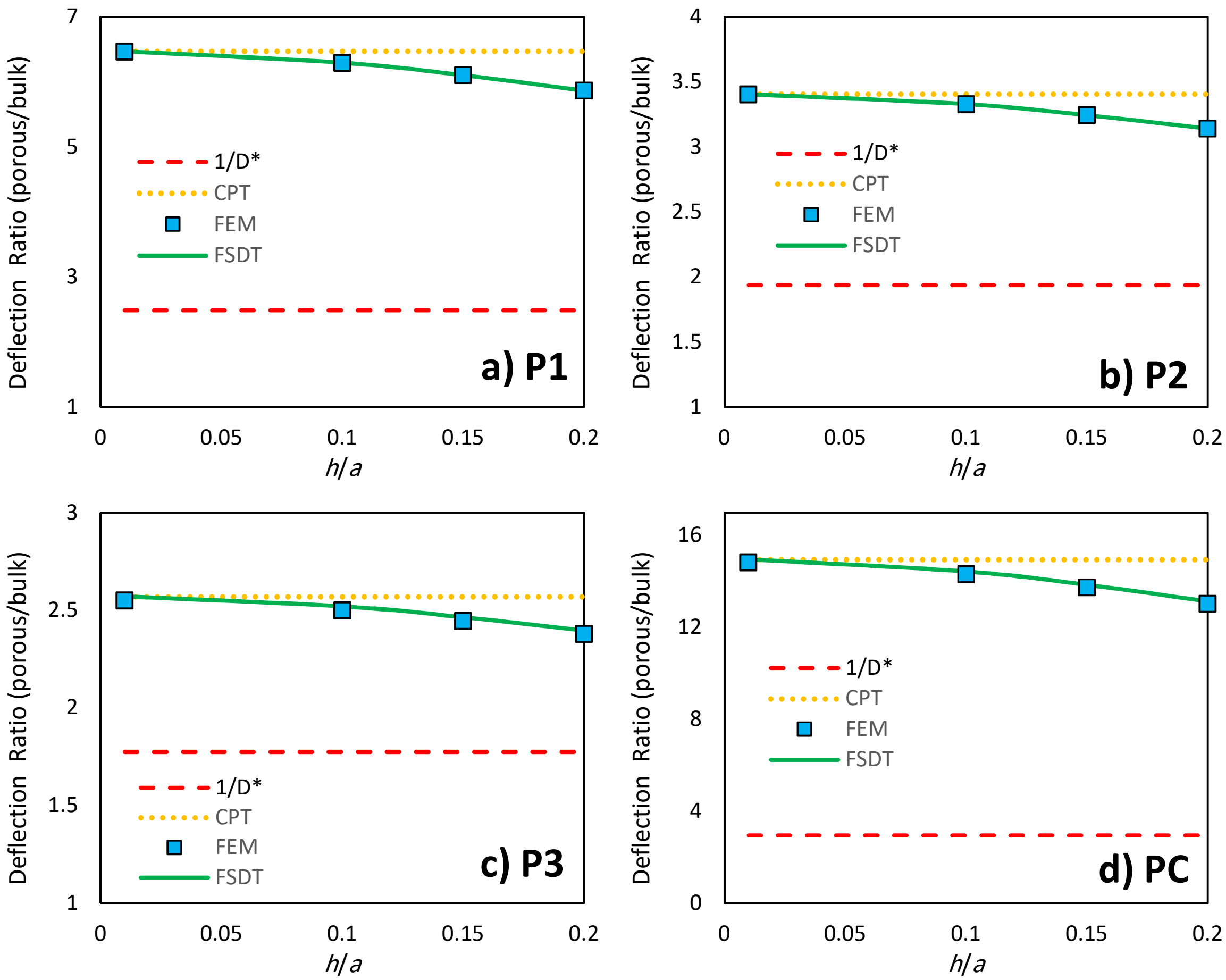

**Fig C3**. Dimensionless deflection for SS-SS-SS-SS square plates versus thickness to side-length ratio under the transverse uniform distributed loading. a) Pyramid Linear P1, b) Pyramid Parabolic P2, c) Pyramid Cubic P3, d) Pyramid Cosine PC.